\documentclass[conference,compsoc]{IEEEtran}
\ifCLASSOPTIONcompsoc
  \usepackage[nocompress]{cite}
\else
  \usepackage{cite}
\fi
\ifCLASSINFOpdf
   \usepackage[pdftex]{graphicx}
\else
\fi
\begin{document}
%
\title{A Survey on Security in Named Data Networking}

\author{\IEEEauthorblockN{Shuoshuo Chen}
\IEEEauthorblockA{University of Virginia\\
Charlottesville, Virginia 22903\\
Email: sc7cq@vieginia.edu}
\and
\IEEEauthorblockN{Fabrice Mizero}
\IEEEauthorblockA{University of Virginia\\
Charlottesville, Virginia 22903\\
Email: fm9ab@virginia.edu}}


%


\maketitle

\begin{abstract}
Over the past three decades, since its invention, the Internet has evolved in both its sheer volume and usage. The Internet's core protocol, Internet Protocol (IP), has proven its usability and effectiveness to support a communication network. However, current Internet usage requires more than a communication network due to a shift in the nature of Internet applications from simple email application to large content producers such as NetFlix, Google, Amazon, etc. Named Data Networking (NDN) is one of the few initiatives/projects addressing the shortcomings of the current Internet architecture and intends to move the Internet toward a content distribution architecture. In this paper, we conduct a brief survey of security topics/problems inherent to the NDN architecture. Specifically, we describe current known problems and propose solutions to major security problems. 
\end{abstract}

%
\IEEEpeerreviewmaketitle

\section{Introduction}
\label{introduction}
According to a most recent forecast by Cisco \cite{index2015forecast}, the content delivery network (e.g. Netflix, Akamai and etc.) has carried about 39\% of the global Internet traffic by 2014. By 2019, this number is projected to grow up to 62\%. Such growth rate leads to one major observation apropos the current usage of the Internet: while originally designed for end-to-end communication, today's Internet is mainly used as a content distribution network. As the capacity and type of applications of the Internet keeps growing, this disparity could put more pressure on the existing architecture, which it was not designed to handle. Adapting the Internet Architecture to its current usage would require complicated fixes and, in some cases, re-engineering key components.

Information-Centric Networking (ICN) \cite{jacobson2006new} is an emerging movement in the Internet architecture revolution proposed to address the mismatch of the current architecture and its usage. ICN proposes new designs aimed at reshaping the Internet into a content distribution architecture. One of those design is Named Data Networking (NDN) \cite{jacobson2009networking,zhang2014named}. NDN is one of the five research projects funded by NSF under its Future Internet Architecture Program, which embodies the concepts in ICN. The basic idea behind NDN is simple: instead of using packets that name the end hosts, name the datagrams themselves. In such a manner, the communication-centric network is turned into an information-centric network, which naturally addresses the content delivery and mobility support challenges.

While the Internet was originally designed as a communication network, the emergence of large content providers such as Google, NetFlix, Amazon, etc. has lead to several changes to the Internet Architecture. Among others, Content Delivery Networks (CDN) service providers use parallel TCP connections \cite{akiyama2015saps} and WAN acceleration \cite{zhang2015using} to disseminate the content closer to consumers. In NDN architecture, the content is virtually cached in every router, thus eliminating the need for CDN services while providing same benefits with in-network memory. Such simple and straight forward solution demonstrates how an architectural change in the Internet design can intrinsically solve problems that would take complicated undertaking otherwise.

Though there are many interesting topics in NDN, we limit this survey within the scope of security. This paper is organized as follows: Section \ref{background} talks about the basic concepts of NDN, current development and its security solution. Section \ref{Advantages and Disadvantages of NDN} describes the difference between IP and NDN with a focus on security, their pros and cons, and open questions in NDN. In section \ref{Possible Solutions}, we propose some possible solutions for the questions. Section \ref{Conclusion} concludes this paper.

\section{Background}
\label{background}
\subsection{Overview}
NDN defines two types of packets: Interest and Data. Interest packets contain the name of a data chunk that the user is looking for. Data packets contain the actual content associated with the specified name. The data retrieval in NDN can be considered as a pull model. A data chunk is fetched when requested by an interest. And the interest should be sent out by a user who needs the data. One major difference in NDN is that network now is treated as an entity that has the information we need. While in IP world, the network is more of a transparent infrastructure that passes information along.

NDN runs the forwarding functionality based on named data chunks. Routers use the Pending Interest Table (PIT) and the Forwarding Information Base (FIB) to decide where packets should be sent. The Content Store (CS) acts as a cache to speed up some of the requests. If an Interest hits in a router's CS, the corresponding data packet will be returned from the local CS. Otherwise, the router forwards the Interest to its next hop based on some pre-computed strategy.

The namespace in NDN is almost unbounded because it assumes the name of a packet can be any string, any length. A current implementation of the namespace is the URL-like semantic hierarchical labels delineated by '/'. Such nomenclature clearly defines the relationship among multiple data chunks, which is the fundamental of the trust model to be described. In addition to the name field and content field, every NDN packet has a signature signed by the data producer that assures the integrity of a packet. While validating a data packet with a given key is trivial, the main challenge lies in determining if a key is authenticate for a packet. To address this problem, a hierarchical key space structures the authentication chain.

\subsection{Security and Privacy in NDN}
Smetters and Jacobson first proposed three properties for a receiver to verify when receiving a data packet: validity, provenance, and relevance \cite{smetters2009securing}. They also gave the idea of self-certifying names such that validity can be verified by running SHA-1 against the content and comparing with its name.

\begin{figure}[!t]
    \centering
    \includegraphics[width=0.4\textwidth]{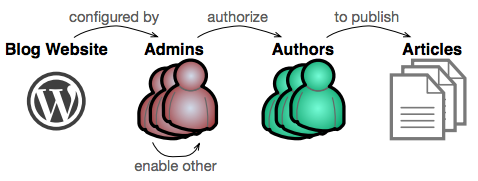}
    \caption{Blog Trust Model \cite{yuschematizing}}
    \label{fig:blog-trust-model}
\end{figure}
The authentication chain in NDN is described in \cite{yuschematizing}. A blog website framework depicts the trust model where the root key signs the administrator key, the administrator key signs the other administrator keys, any of the administrator key signs the author key, and the author key signs corresponding articles. Figure \ref{fig:blog-trust-model} illustrates the visual relationship in this blog website example. The authentication across different namespaces could be very complicated. Thus, it is necessary to automate the authentication process. A Trust schema is designed to allow such automation. By using a semantic rule pattern in names, a consumer can analyze the names and parse them like regular expressions to figure out the relationship. 

To minimize the threat by a compromised consumer or producer, the keys \cite{yuconfidentiality} have a lifetime property. Each key is created based on the minimum privilege principle. The less privilege a key has, the shorter its lifetime. In such a way, even if a user is compromised, he/she can only use the key for a limited amount of time, which limits the potential damage.

\begin{figure}[!ht]
    \centering
    \includegraphics[width=0.45\textwidth]{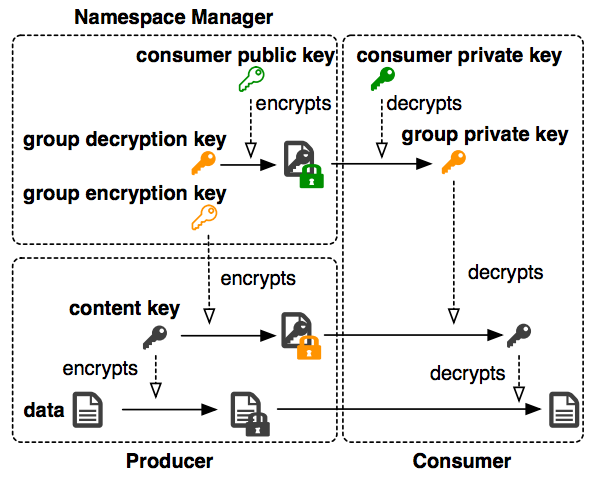}
    \caption{Named-based Access Control \cite{yuconfidentiality}}
    \label{fig:access-control}
\end{figure}

Figure \ref{fig:access-control} illustrates how access control is designed in name-based key distribution system. A group contains some producers and some consumers, producing or consuming data of particular interests. The Namespace manager acts as the Certificate Authority (CA) to give out root certificates. In a typical scenario, the namespace manager holds private keys from consumers in the group. Then it creates a public/private key pair for the group. Since the public key can be transferred in plain text, it is directly given to the producer without encryption. The private key is encrypted by each consumer's public key respectively and sent to each consumer. A consumer can later decrypt using its local private key to get the group private key. Once the producer has the group public key and the consumer has the group private key, the user access control is done. Next, the producer needs to set up the content access control. The producer uses the group public key to encrypt a symmetric content key and sends it to its consumers. The consumer can decrypt it using its group private key. Now, both ends have the symmetric key. In this way, the producer can encrypt the content using the content key. And consumer can decrypt content using the same key.

Another topic in security is the Denial-of-Serive (DoS) attack. In IP architecture, despite several research works, no efficient, and accurate solution has been found so far. Most DoS require monitoring traffic using NetFlow and conducting offline analsysis of traffic flows. Such methods are slow relatively to the speeds of sophisticated attacks, such as Distributed DoS (DDoS). Unlike the current Internet architecture, NDN has natural advantages in preventing DoS attacks. Since NDN addressing is based on content names rather than destination (end hosts) addresses, it is almost impossible to  direct all requests to a designated spot as is required in a typical DoS attack. The CS in NDN routers plays a major role in mitigating DoS attacks by reducing the load on data/content producers. However, researchers soon realized there are similar attacks in NDN. Despite NDN's ability to effortlessly combat DoS attacks, researchers have found its vulnerabilities in other type of attacks such as Interest flooding attack, as explained in \cite{afanasyev2013interest}. Basically, by excessively requesting data within a certain namespace, a router could soon use up its memory and PIT entries, thus leading to denial of service for other users served by the same router \cite{compagno2013poseidon}.

Cache pollution is also a potential category of attack in NDN \cite{conti2013lightweight}. An adversary could leverage a Bot-net to fill up a router's local CS with rarely requested content in order to downgrade the service quality for other users.

Cache privacy problem proposed in \cite{acs2013cache} is a more targeted threat on user. An adversary can decide whether a user has requested some particular data in a following way. The adversary first requests for the content $C$ that a user could have requested and logs the delay $d$. Then it requests another data chunk $C'$ twice that user must have not requested. Thus, for the first time, delay will be $d_1$ and the second time delay will be $d_2$, where $d_1$ \textgreater $d_2$, since the second request hits the cache. The adversary can therefore know whether $C$ is cached by comparing $d$ and $d_2$.

\section{Advantages and Disadvantages of NDN}
\label{Advantages and Disadvantages of NDN}

\subsection{Comparison between IP and NDN}
From a high level perspective, NDN architecture has a similar structure as IP architecture. Both share the structure of hourglass with narrow waist. In IP architecture, the narrow waist is IP packets serving as a common layer across all types of underlying networks and upper applications. In NDN, the narrow waist is NDN packets serving as the common layer. Both IP and NDN follow Clark's end-to-end principle \cite{saltzer1984end}. The routers do not care what kind of applications is utilizing the packets, routers only take care of the packets forwarding. And both architectures appear as layered structures. Either IP or NDN stands as an outstanding abstract layer that provides service for all the upper layer applications.

The most significant difference between IP and NDN is the change of service type. The traditional scheme of delivering data to a destination now becomes a new scheme of fetching data associated with a particular name. IP routers are stateless, they do not store any information about the packets they have processed. While NDN routers are stateful, requested packets are cached in CS not only for acceleration but also affect the forwarding strategy a router makes. In terms of the architectural design, there are several pros and cons existing between IP and NDN. We only discuss from a security perspective.

Network layer security is a significant improvement in NDN that IP does not originally support. Through the use of certificates, every data packet in NDN is signed at the producer and can be verified at the consumer via keys as explained earlier. Per-packet encryption add a security guarantee in network layer. In IP architecture, security is implemented through end-to-end channel by upper layer protocols like (Secure Socket Layers) SSL. These protocols only provides limited security support for applications as they establish a secure communication channel rather than securing the contents. However, several major protocols in IP such as BGP, OSPF, ICMP and etc. are not encrypted. These unsecured protocols could be utilized to launch attacks. By requiring a signature on every packet, NDN  effectively addresses the integrity and provenance of a packet.

An advantage of IP over NDN is its many years of engineering evaluation (over 30 years) by billions of users, which has ultimately proved its effectiveness and usability. In contrast, NDN is a relatively new concepts in its experimental phase. As of the time of this writing, for practical purposes, NDN is implemented on top of IP on its major cross-continent testbed. Without the actual development of router hardware (ASICs) to handle the many components of NDN and further public testing, it is unfair to compare IP and NDN despite many foreseen theoretical advantages the latter guarantees.  As with many protocols, only extensive public use can reveal major shortcomings. In the case of TCP, congestion problems were found after DARPAnet was put into use. In addition, the more time a protocol spends in non-experimental use, the more sophisticated cyber attacks become. NDN must also pass this test of time.

\subsection{Open Questions}
As mentioned previously, there are several key concerns in NDN security. In this section, we discuss them in the following order: (i) key distribution; (ii) key revocation; (iii) Interest flood; and (iv) cache pollution.

NDN key distribution scheme follows the trust model discussed in the previous section. However, a few problems are also exposed in such a trust model. First, since the actual content is encrypted by symmetric key, an attacker can use brute force to crack the encrypted content. Current computational power can easily crack a 56-bit DES encryption system in minutes.  Increasing the length of the keys can prevent such possibility. However, a long keys  would lead to overhead at the router, which could cause low throughput and/or unexpectedly high delays. Moreover, the major problem lies in the namespace manager. 
The namespace manager needs to store every single public key of a consumer. Considering a large group of consumers, the space to store these keys could be very large and the key management could also be of low efficiency. Such a centralized manager is similar to the certificate authority (CA) in IP world, which is prone to leaking all the keys when compromised. Besides, the trust schema as interpreted as a chain of security rules is dependent on the namespace design. Since NDN leaves the choice of data names to the application, there needs to be some forcible rules to prevent attacks by maliciously constructing the name. In a simple trust schema, the risk of failing an authentication is high when rules are too simple. A robust way is to apply multi-level chained authentication. However, this inevitably lowers the efficiency because the chain needs to be walked through every time.

One possible way of key revocation is by negative statements. A user can explicitly announce the revocation of his key to the public. Thereafter, other users in the network can thus react by updating their local configuration respectively \cite{yupublic}. However, this negative statement is not guaranteed to be received by everyone. A malicious attacker could: fill a packet with harmful content, respond to Interests and sign the packet with revoked key. If the attacker manages to prevent a target user from receiving the revocation statement, this attack could work. On the other hand, if a key is compromised, all the keys generated by this compromised key need to be revoked correspondingly. Due to the multi-layered hierarchy in key privileges, there is no guarantee that no missed key can still exist and be used.

A relatively effective way is to make the origin owner of the key to revoke it. But sometimes it is impossible because the user might still need the key. Even it is possible, the problem of reachability still exists. And the ultimate problem is how to limit the number of affected users if the key is compromised.

Though NDN can prevent the regular type of DoS attack pretty well, there is an equivalent type of attack in NDN realm, Interest flood \cite{gasti2013and}. An adversary cannot target a specific host or router but it can target a namespace. If the name he requests is (i) unique and a particular host is the lone producer for its corresponding namespace, (ii) can respond to the request, (iii) and provides the corresponding data exclusively, the adversary can make use of this name to plan an attack. All the interests passing through the intermediate routers will be forwarded towards that host. In such a way, the adversary needs to make sure that the interests are passed along by the routers and every packet he requests has a unique name that only resides in that host. Then he needs to keep sending out interests towards the target. As a consequence, the PIT could effectively be depleted by such an excessive amount of interests, which eventually leads to disruption of service.

As for cache pollution, the ultimate goal is to destroy cache localization and deteriorate the link utilization. An adversary can compromise a small set of consumers and use them to issue interests. By carefully requesting for unpopular content, the adversary could force the router to cache those unpopular content in order to deplete its available local CS space. Once the CS is filled with unpopular content from an attacker, other consumers would need to fetch data from further places. This essentially exacerbates link utilization and throughput and affects all the connected users.

\section{Possible Solutions}
\label{Possible Solutions}
\textbf{Centralized namespace management} A centralized namespace manager may lead to several security threats that endanger a wide range of users. As a single point of failure, it also inherits the shortcomings any centralized system. We propose a distributed namespace controller for several reasons, namely, risk containment, and efficiency. For reliability purposes, it is crucial to have some sort of redundancy in any critical part of a big system such as a big network. Such redundancy prevents the system from completely shutting off in case of a failure. For risk containment purposes, potential threats (such as in cases where a key is compromised) can be discovered and contained at the domain level. A distributed namespace manager is also more efficient. Syncing across few managers is more efficient and effective compared to synching across all routers.

\textbf{Interest flood} Two possible solutions can help mitigate the negative effects of Interest flood: interest load monitoring and policing. On a router level, interest traffic can be monitored in time and analyzed on the fly. To reach near real-time performance, sampling techniques, such as those featured in NetFlow, can be used. In the event where the interest load for a particular namespace reaches a threshold, interests may be dropped selectively. Such policing techniques increases the probability that some portion of the bandwith is always left for non-malicious interest in the same namespace.

\textbf{Cache pollution}
To prevent the problem of cache pollution, we propose using a split content store at the router level. A single CS can be implemented as 2 separate memory slots. By monitoring cache occupancy in time, unpopular cache occupants can be migrated to the smaller slot of the CS. For effectiveness reasons, such migrations can only be performed at regular intervals, $\tau$. After every $\tau$ units of time, a ranking/sorting of cache occupants by the number of hits/access determines the least recently accessed, which are then migrated to the smaller CS slot. Entries in the CS can also be timed to allow eviction of stale content and free the space. Such popularity-based allocations would ensure that popular content is available and close to consumers, thus allowing for a high performance network.
\section{Conclusion}
\label{Conclusion}
Ever since Van Jacobson's seminal talk in 2006, ICN community has made significant strides to making ICN designs a reality. The NDN architecture, in particular, has attracted a lot of researchers and computer scientists. As a result working prototypes of NDN architecture and a cross-continent testbed have made significant expansion possible through research. Theoretically, NDN will solve several problems that IP has been unable to address efficiently over its several decades of existence. In security, NDN addresses security challenges such as DoS attacks naturally and in-router data caching provides more efficient CDN service. However, NDN brings its new security challenges/concerns inherent to its architecture such as interest flooding, cache poisoning, key management problems, that are more or less significant depending on the nature or sensitivity of data. In this paper, we proposed potential solutions to 3 major NDN security problems. We propose three solutions, namely, distributed namespace management, interest load monitoring and policing, and a two-way split Content Store, to alleviate and in same cases, eliminate security threats posed by centralized namespace management, interest flood, and cache pollution, respectively. 



%

\bibliographystyle{IEEEtran}
\bibliography{ref}

\end{document}